\documentclass[copyright,creativecommons]{eptcs}
\usepackage{underscore}
\usepackage{xspace}
\usepackage[T1]{fontenc}
\usepackage{amssymb}
\usepackage{enumitem}
\usepackage{textcomp}
\usepackage{alltt}
\usepackage{xcolor}
\usepackage{graphicx}
\usepackage{multirow}

\newcommand{\acl}[1]{\texttt{#1}}
\newcommand{\iacl}[1]{\texttt{\textit{#1}}}

\newcommand{\sfn}[1]{$\iacl{fn}\!_{#1}$}
\newcommand{\svar}[1]{$\iacl{var}\!_{#1}$}
\newcommand{\sarg}[1]{$\iacl{arg}\!_{#1}$}

\newcommand{\java}[1]{\texttt{#1}}
\newcommand{\ijava}[1]{\texttt{\textit{#1}}}
\definecolor{commentgray}{gray}{0.4}
\newcommand{\cjava}[1]{\texttt{\textcolor{commentgray}{#1}}}
\newenvironment{javablock}{\begin{alltt}}{\end{alltt}}

\newcommand{\centertilde}{\raisebox{0.5ex}{\texttildelow}}

\newcommand{\secref}[1]{Section~\ref{#1}}
\newcommand{\secrefII}[2]{Sections~\ref{#1}~and~\ref{#2}}
\newcommand{\figref}[1]{Figure~\ref{#1}}
\newcommand{\tabref}[1]{Table~\ref{#1}}

\newcommand{\tabrefIII}[3]{Tables~\ref{#1},~\ref{#2},~and~\ref{#3}}
\newcommand{\footref}[1]{Footnote~\ref{#1}}

\newcommand{\acldoc}[2]
 {\cite[\href{#2}{\texttt{:doc} \texttt{#1}}]{acl2-manual}}

\begin{document}


\title{A Simple Java Code Generator for ACL2 \\
       Based on a Deep Embedding of ACL2 in Java}

\author{Alessandro Coglio
        \institute{Kestrel Institute \\ \url{http://www.kestrel.edu}}}

\def\titlerunning{Java Code Generator and Deep Embedding}
\def\authorrunning{Alessandro Coglio}

\maketitle

\begin{abstract}
AIJ (\textbf{A}CL2 \textbf{I}n \textbf{J}ava) is
a deep embedding in Java
of an executable, side-effect-free, non-stobj-accessing subset
of the ACL2 language without guards.
ATJ (\textbf{A}CL2 \textbf{T}o \textbf{J}ava) is
a simple Java code generator
that turns ACL2 functions into AIJ representations
that are evaluated by the AIJ interpreter.
AIJ and ATJ enable possibly verified ACL2 code
to run as, and interoperate with, Java code,
without much of the ACL2 framework or any of the Lisp runtime.
The current speed of the resulting Java code
may be adequate to some applications.
\end{abstract}


\section{Motivation and Contributions}
\label{sec:motiv}

A benefit of writing code in a theorem prover
is the ability to prove properties about it,
such as the satisfaction of requirements specifications.
A facility to generate code in one or more programming languages
from an executable subset of the prover's logical language
enables the possibly verified code to run as, and interoperate with,
code written in those programming languages.
Assuming the correctness of code generation
(whose verification is a separable problem, akin to compilation verification)
the properties proved about the original code carry over to the generated code.

The ACL2 theorem provers's tight integration with the underlying Lisp platform
enables the executable subset of the ACL2 logical language
to run readily and efficiently as Lisp,
without the need for explicit code generation facilities.
Nonetheless, some situations may call for
running ACL2 code in other programming languages:
specifically, when the ACL2 code must interoperate
with external code in those programming languages
in a more integrated and efficient way than afforded
by inter-language communication via foreign function interfaces \cite{cffi,jni},
or by inter-process communication with the ACL2/Lisp runtime
via mechanisms like the ACL2 Bridge
\acldoc{bridge}{http://www.cs.utexas.edu/users/moore/acl2/manuals/current/manual/?topic=ACL2____BRIDGE}.
Using Lisp implementations
written in the target programming languages \cite{abcl}
involves not only porting ACL2 to them,
but also including much more runtime code
than necessary for the target applications.
Compilers from Lisp to the target programming languages
may need changes or wrappers,
because executable ACL2 is not quite a subset of Lisp;
furthermore, the ability to compile non-ACL2 Lisp code
is an unnecessary complication as far as ACL2 compilation is concerned,
making potential verification harder.

The work described in this paper contributes
to the goal of running ACL2 code in other programming languages
in the integrated manner described above:
\begin{itemize}[nosep]
\item
ATJ (\textbf{A}CL2 \textbf{T}o \textbf{J}ava)
is a Java code generator for ACL2.
ATJ translates
executable, side-effect-free, non-stobj-accessing ACL2 functions,
without their guards,
into Java.
It does so in a simple way,
by turning the functions into deeply embedded Java representations
that are executed by an ACL2 evaluator written in Java.
\item
AIJ (\textbf{A}CL2 \textbf{I}n \textbf{J}ava)
is a deep embedding in Java
of an executable, side-effect-free, non-stobj-accessing subset
of the ACL2 language without guards.
AIJ consists of
(i) a Java representation of the ACL2 values, terms, and environment,
(ii) a Java implementation of the ACL2 primitive functions, and
(iii) an ACL2 evaluator written in Java.
AIJ executes the deeply embedded Java representations of ACL2 functions
generated by ATJ.
AIJ is of independent interest and can be used without ATJ.
\end{itemize}
The ACL2 language subset supported by ATJ and AIJ includes
all the values,
all the primitive functions,
and many functions with raw Lisp code---%
see \secref{sec:evalsem} for details on these two kinds of functions.

The initial implementation of AIJ
favored assurance over efficiency:
it was quite simple,
to reduce the chance of errors
and facilitate its potential verification,
but it was also quite slow.
The careful introduction of a few optimizations,
which do not significantly complicate the code
but provide large speed-ups,
makes the speed of the current implementation
arguably adequate for some applications;
see \secref{sec:tests}.
Furthermore, the code is amenable to additional planned optimizations;
see \secref{sec:future}.


\section{Background: The Evaluation Semantics of ACL2}
\label{sec:evalsem}

ACL2 has a precisely defined logical semantics \cite{acl2-logic},
expressed in terms of syntax, axioms, and inference rules,
similarly to logic textbooks and other theorem provers.
This logical semantics applies to logic-mode functions,
not program-mode functions.
Guards are not part of the logic,
but engender proof obligations in the logic
when guard verification is attempted.

ACL2 also has a documented evaluation semantics
\acldoc{evaluation}{http://www.cs.utexas.edu/users/moore/acl2/manuals/current/manual/?topic=ACL2____EVALUATION},
which could be formalized
in terms of syntax, values, states, steps, errors, etc.,
as is customary for programming languages.
This evaluation semantics applies
to both logic-mode and program-mode functions.
Guards affect the evaluation semantics,
based on guard-checking settings.
Even non-executable functions
(e.g.\ introduced via \acl{defchoose} or \acl{defun-nx})
degenerately have an evaluation semantics,
because they do yield error results when called;
however, the following discussion focuses on executable functions.

Most logic-mode functions have definitions
that specify both their logical and their evaluation semantics:
for the former, the definitions are logically conservative axioms;
for the latter, the definitions provide ``instructions''
for evaluating calls of the function.
For a defined logic-mode function,
the relationship between the two semantics is that,
roughly speaking,
evaluating a call of the function yields, in a finite number of steps,
the unique result value that, with the argument values,
satisfies the function's defining axiom---%
the actual relationship is slightly more complicated,
as it may involve guard checking.

The primitive functions \acldoc{primitive}{http://www.cs.utexas.edu/users/moore/acl2/manuals/current/manual/?topic=ACL2____PRIMITIVE}
are in logic mode and have no definitions;
they are all built-in.
Examples are \acl{equal}, \acl{if}, \acl{cons}, \acl{car}, and \acl{binary-{}+}.
Their logical semantics is specified by axioms of the ACL2 logic.
Their evaluation semantics is specified by raw Lisp code
(under the hood).
The relationship between the two semantics is as in the above paragraph,
with the slight complication that \acl{pkg-witness} and \acl{pkg-imports}
yield error results when called on unknown package names.
The evaluation of calls of \acl{if} is non-strict, as is customary.

Most program-mode functions have definitions
that specify their evaluation semantics,
similarly to the non-primitive logic-mode functions discussed above.
Their definitions specify no logical semantics.

The logic-mode functions
listed in the global variable \acl{logic-fns-with-raw-code}
have a logical semantics specified by their ACL2 definitions,
but an evaluation semantics specified by raw Lisp code.
(They are disjoint from the primitive functions, which have no definitions.)
For some of these functions, e.g.\ \acl{len},
the raw Lisp code just makes them run faster
but is otherwise functionally equivalent to the ACL2 definitions.
Others have side effects,
carried out by their raw Lisp code
but not reflected in their ACL2 definitions.
For example, \acl{hard-error} prints a message on the screen
and immediately terminates execution, unwinding the call stack.
As another example, \acl{fmt-to-comment-window}
prints a message on the screen, returning \acl{nil} and continuing execution.
But the ACL2 definitions of both of these example functions
just return \acl{nil}.

The program-mode functions
listed in the global variable \acl{program-fns-with-raw-code}
have an evaluation semantics specified by raw Lisp code.
Their ACL2 definitions appear to have no actual use.

Since stobjs
\acldoc{stobj}{http://www.cs.utexas.edu/users/moore/acl2/manuals/current/manual/?topic=ACL2____STOBJ}
are destructively updated,
functions that manipulate stobjs may have side effects as well---%
namely, the destructive updates.
Because of single-threadedness,
these side effects are invisible
in the end-to-end input/output evaluation of these functions;
however, they may be visible in some formulations of the evaluation semantics,
such as ones that comprehend interrupts,
for which updating a record field in place involves different steps
than constructing a new record value with a changed field.
The built-in \acl{state} stobj \acldoc{state}{http://www.cs.utexas.edu/users/moore/acl2/manuals/current/manual/?topic=ACL2____STATE}
is ``linked'' to external entities,
e.g.\ the file system of the underlying machine.
Thus, functions that manipulate \acl{state}
may have side effects on these external entities.
For example, \acl{princ\$} (a member of \acl{logic-fns-with-raw-code})
writes to the stream associated with the output channel argument,
and affects the file system.

The fact that the side effects of the evaluation semantics
are not reflected in the logical semantics
is a design choice
that makes the language more practical for programming
while retaining the ability to prove theorems.
But when generating Java or other code,
these side effects should be taken into consideration:
for instance,
turning \acl{hard-error} and \acl{fmt-to-comment-window}
into Java code that returns (a representation of) \acl{nil},
would be incorrect or at least undesired.
As an aside,
a similar issue applies to the use of APT transformations \cite{apt}:
for instance,
using the \acl{simplify} transformation \cite{simplify}
to turn calls of \acl{hard-error} into \acl{nil},
while logically correct and within \acl{simplify}'s stipulations,
may be undesired or unexpected.


\section{AIJ: The Deep Embedding}
\label{sec:aij}

AIJ is a Java package whose public classes and methods
provide an API to
(i) build and unbuild representations of ACL2 values,%
\footnote{When talking about AIJ,
this paper calls `build' and `unbuild'
what is often called `construct' and `destruct' in functional programming,
because in object-oriented programming the latter terms
may imply object allocation and deallocation,
which is not necessarily what the AIJ API does.}
(ii) build representations
of ACL2 terms and of an ACL2 environment,
and (iii) evaluate calls of ACL2 primitive and defined functions,
without checking guards.
By construction, the ACL2 code represented and evaluated by AIJ
is executable, has no side effects, does not access stobjs, and has no guards.

AIJ consists of a few thousand lines of Java code
(including blank and comment lines),
thoroughly documented with Javadoc comments.
The implementation takes advantage of object-oriented features
like encapsulation, polymorphism, and dynamic dispatch.

\begin{figure}
\centering
\includegraphics[width=\textwidth]{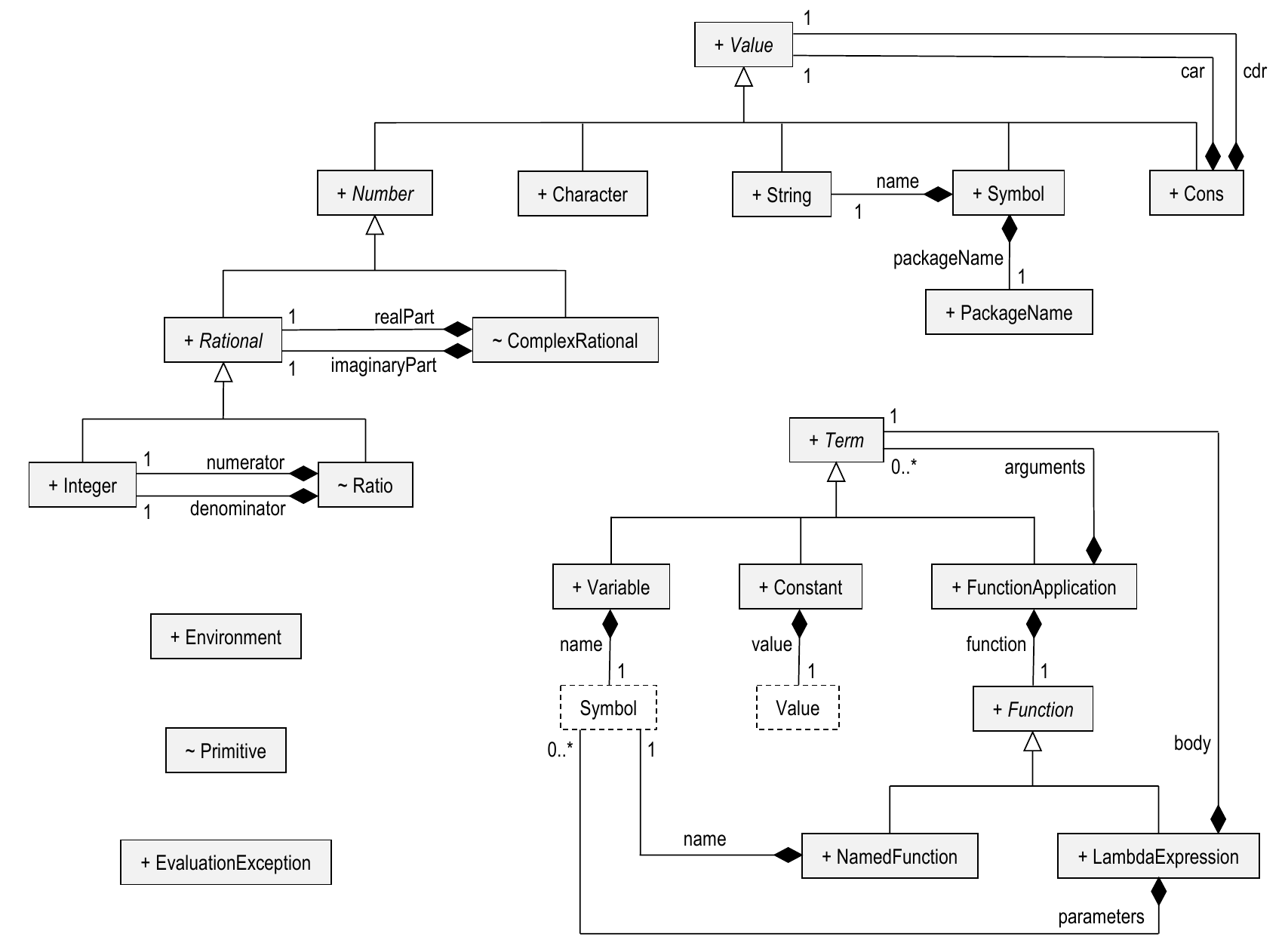}
\caption{\label{fig:uml}
Simplified UML class diagram for AIJ.}
\end{figure}

The Java classes that form AIJ
are shown in the simplified UML class diagram
in \figref{fig:uml}
and described in the following subsections.
Each class is depicted as a box containing its name.%
\footnote{In AIJ's actual code,
each class name is prefixed with `\java{Acl2}' (e.g.\ \java{Acl2Value}),
so that external code can reference these classes unambiguously
without AIJ's package name \java{edu.kestrel.acl2.aij}.
This paper omits the prefix for brevity,
and uses fully qualified names for the Java standard classes
to avoid ambiguities,
e.g.\ \java{java.lang.String} is the Java standard string class,
as distinguished from \java{String} in \figref{fig:uml}.}
Abstract classes have italicized names.
Public classes have names preceded by \java{+},
while package-private classes have names preceded by \java{\centertilde}.
Inheritance (`is a') relationships
are indicated by lines with hollow triangular tips.
Composition (`part of') relationships
are indicated by lines with solid rhomboidal tips,
annotated with
the names of the containing class instances' fields
that store the contained class instances,
and with the multiplicity of the contained instances
for each containing instance
(`0..*' means `zero or more').
The dashed boxes are just replicas to avoid clutter.
This UML class diagram is simplified because
the class boxes do not contain fields and methods.

\subsection{Values}
\label{sec:values}

\begin{figure}
\centering
\includegraphics[width=0.7\textwidth]{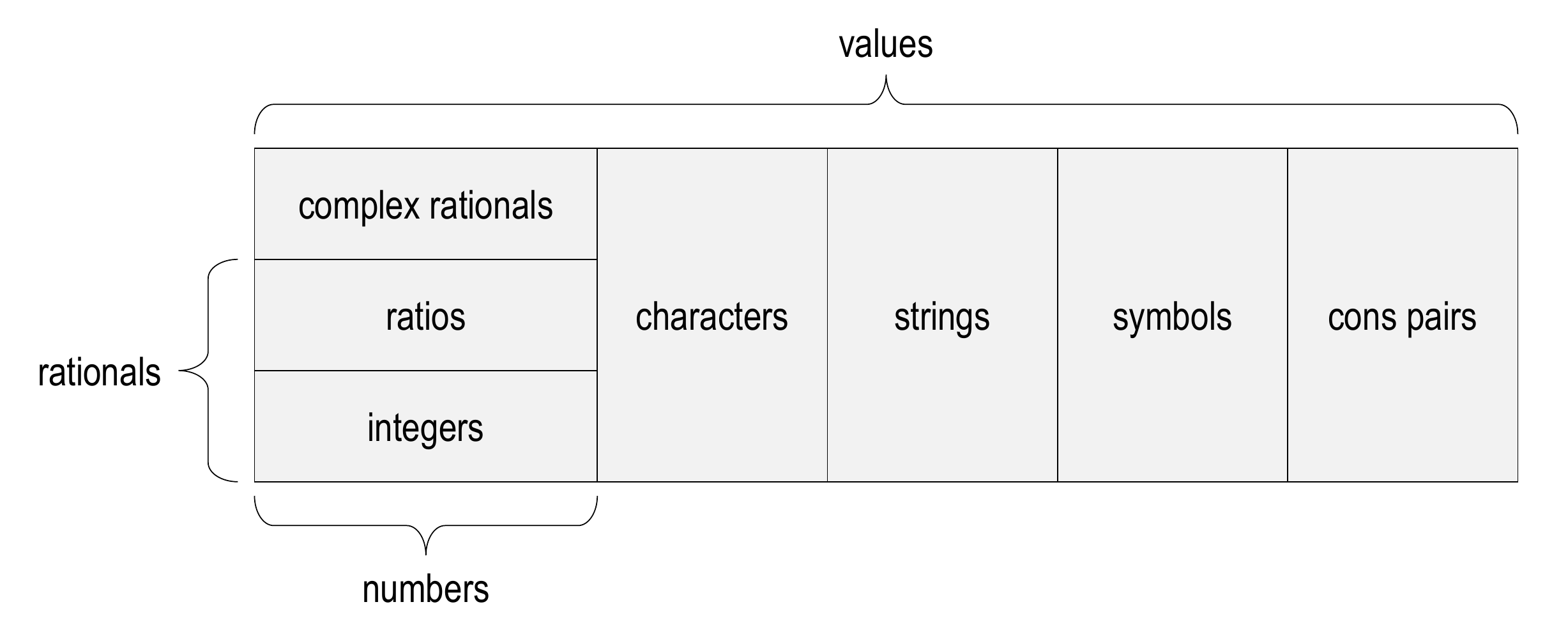}
\caption{\label{fig:values}
Values of the ACL2 evaluation semantics.}
\end{figure}

The set of values of the ACL2 evaluation semantics
is the union of the sets depicted in \figref{fig:values}:
(i) integers, recognized by \acl{integerp};
(ii) ratios, i.e.\ rationals that are not integers,
with no built-in recognizer;%
\footnote{The term `ratio' is used in the Common Lisp specification
\cite[Section 2.1.2]{cltl2}.}
(iii) complex rationals, recognized by \acl{complex-rationalp};
(iv) characters, recognized by \acl{characterp};
(v) strings, recognized by \acl{stringp};
(vi) symbols, recognized by \acl{symbolp}; and
(vii) \acl{cons} pairs, recognized by \acl{consp}.
Integers and ratios form the rationals, recognized by \acl{rationalp}.
Rationals and complex rationals form the Gaussian rationals,
which are all the numbers in ACL2,
recognized by \acl{acl2-numberp}.%
\footnote{This discussion does not apply to ACL2(r).}
The logical semantics of ACL2 allows additional values called `bad atoms',
and consequently \acl{cons} pairs that may contain them directly or indirectly;
however, such values cannot be constructed in evaluation.

AIJ represents ACL2 values
as immutable objects of \java{Value} and its subclasses in \figref{fig:uml}.
Each such class corresponds to a set in \figref{fig:values}.
The subset relationships in \figref{fig:values}
match the inheritance relationships in \figref{fig:uml}.
The sets of values that are unions of other sets of values
correspond to abstract classes;
the other sets correspond to concrete classes.
All these classes are public,
except for the package-private ones for ratios and complex rationals:
ratios and complex rationals are built indirectly via AIJ's API,
by building
rationals that are not integers and numbers that are not rationals.

The information about the represented ACL2 values
is stored in fields of the non-abstract classes.
\java{Integer} stores
the numeric value as a \java{java.math.BigInteger}.
\java{Ratio} stores
the numerator and denominator as \java{Integer}s,
in reduced form
(i.e.\ their greatest common divisor is 1
and the denominator is greater than 1).
\java{ComplexRational} stores
the real and imaginary parts as \java{Rational}s.
\java{Character} stores
the 8-bit code of the character as a \java{char} below 256.
\java{String} stores
the codes and order of the characters as a \java{java.lang.String}
whose \java{char}s are all below 256.
\java{Symbol} stores
the symbol's package name as a \java{PackageName}
(a wrapper of \java{java.lang.String}
that enforces the ACL2 constraints on package names)
and the symbol's name as a \java{String}.
\java{Cons} stores the component \java{Value}s.
All these fields are private,
thus encapsulating the internal representation choices
and enabling their localized modification.
ACL2 numbers, strings, and symbols have no preset limits,
but the underlying Lisp runtime may run out of memory.
Their Java representations (e.g.\ \java{java.math.BigInteger})
have very large limits,
whose exceedance could be regarded as running out of memory.
If needed, the Java representations could be changed
to overcome the current limits
(e.g.\ by using lists of \java{java.math.BigInteger}s).

The public classes for ACL2 values and package names
provide public static factory methods to build objects of these classes.
For example, \java{Character.make(char)}
returns a \java{Character} with the supplied argument as code,
throwing an exception if the argument is above 255.
As another example, \java{Cons.make(Value,Value)}
returns a \java{Cons} with the supplied arguments as components.
Some classes provide overloaded variants,
e.g.\ \java{Integer.make(int)} and \java{Integer.make(java.math.BigInteger)}.
All these classes provide no public Java constructors,
thus encapsulating the details of object creation and re-use,
which is essentially transparent to external code
because these objects are immutable.

The public classes for ACL2 values provide public instance getter methods
to unbuild (i.e.\ extract information from) objects of these classes.
For example, \java{Character.getJavaChar()}
returns the code of the character
as a \java{char} that is always below 256.
As another example, \java{Cons.getCar()} and \java{Cons.getCdr()}
return the component \java{Value}s of the \acl{cons} pair.
Some classes provide variants,
e.g.\ \java{Integer.getJavaInt()}
(which throws an exception if the integer does not fit in an \java{int})
and \java{Integer.getJavaBigInteger()}.


\subsection{Terms}
\label{sec:terms}

ACL2 translates the terms supplied by the user,
which may include macros and named constants,
into a restricted internal form,
in which macros and named constants are expanded
\acldoc{term}{http://www.cs.utexas.edu/users/moore/acl2/manuals/current/manual/?topic=ACL2____TERM}.
In the rest of this paper,
`term' means `translated term',
i.e.\ a term in the restricted internal form.

The set of ACL2 terms consists of
(i) variables,
which are symbols,
(ii) quoted constants,
which are lists \acl{(quote \iacl{value})} where \iacl{value} is a value,
and (iii) function applications,
which are lists \acl{(\iacl{fn} \sarg{1} ... \sarg{n})}
where \iacl{fn} is a function
and \sarg{1}, ..., \sarg{n} are zero or more terms.
A function \iacl{fn} used in a term is
(i) a named function,
which is a symbol,
or (ii) a lambda expression,
which is a list \acl{(lambda (\svar{1} ... \svar{m}) \iacl{body})}
where \svar{1}, ..., \svar{m} are zero or more symbols
and \iacl{body} is a term,
whose free variables are all among \svar{1}, ..., \svar{m}
(i.e.\ lambda expressions are always closed).

AIJ represents ACL2 terms in a manner similar to ACL2 values,
as immutable objects of \java{Term} and its subclasses in \figref{fig:uml};
functions are represented
as immutable objects of \java{Function} and its subclasses in \figref{fig:uml}.
The superclasses are abstract, while the subclasses are concrete.
All these classes are public.

The information about the represented ACL2 terms
is stored in private fields of the non-abstract classes.
\java{Variable} and \java{NamedFunction} are wrappers of \java{Symbol}.
\java{Constant} is a wrapper of \java{Value}.%
\footnote{These wrappers place \java{Symbol}s and \java{Value}s
into the class hierarchy of \java{Term} and \java{Function},
given that Java does not support multiple class inheritance.
For instance, \java{Symbol} could not be
both a subclass of \java{Value} and a subclass of \java{Term}.}
\java{FunctionApplication} stores
a \java{Function} and an array of zero or more \java{Term}s.
\java{LambdaExpression} stores
an array of zero or more \java{Variable}s and a \java{Term}.

The non-abstract classes for ACL2 terms (and functions) provide
public static factory methods to build objects of these classes,
but no public Java constructors,
similarly to the classes for ACL2 values.


\subsection{Environment}
\label{sec:environment}

ACL2 terms are evaluated in an environment that includes
function definitions, package definitions, etc.
AIJ stores information about part of this environment
in \java{Environment} in \figref{fig:uml}.
Since there is just one environment at a time in ACL2,
this class has no instances and only static fields and methods.

An ACL2 function definition consists of several pieces of information,
of which AIJ only stores
(i) the name, which is a symbol,
(ii) the parameters, which are zero or more symbols,
and (iii) the body, which is a term.
\java{Environment} stores function definitions in a private static field,
as a \java{java.util.Map}
from \java{Symbol}s for the functions' names
to \java{LambdaExpression}s for the functions' parameters and bodies.
The public static method
\java{Environment.addFunctionDef(Symbol,Symbol[],Term)}
adds a function definition to the map.

An ACL2 package definition
associates a list of imported symbols to a package name.
\java{Environment} stores package definitions in a private static field,
as a \java{java.util.Map}
from \java{PackageName}s for the packages' names
to \java{java.util.List}s of \java{Symbol}s for the packages' import lists.
The public static method
\java{Environment.addPackageDef(PackageName,List<Symbol>)}
adds a package definition to the map.
AIJ uses this field to implement the primitive function \acl{pkg-imports}.
AIJ also uses information derived from this field to implement
the overloaded factory methods \java{Symbol.make} that build symbols:
for instance, \java{Symbol.make("ACL2","CONS")} returns a \java{Symbol}
with name \acl{"CONS"} and package name \acl{"COMMON-LISP"},
not package name \acl{"ACL2"},
because \acl{"ACL2"} imports \acl{cons} from \acl{"COMMON-LISP"};
the call \java{Symbol.make("ACL2","CONS")} is the Java equivalent
of the ACL2 symbol notation \acl{acl2::cons}.

\java{Environment} also stores
the value of the ACL2 constant \acl{*pkg-witness-name*}
in a private static field, as a \java{java.lang.String}.
This field may be set, at most once (otherwise an exception is thrown),
via the public static method
\acl{Environment.setPackageWitnessName(java.lang.String)}.
AIJ uses this field to implement the primitive function \acl{pkg-witness}.


\subsection{Primitive Functions}
\label{sec:primitives}

Since the ACL2 primitive functions have no definitions,
AIJ cannot evaluate their calls via their bodies
as described in \secref{sec:evaluation}.
AIJ implements these functions ``natively'' in Java,
in the package-private class \java{Primitive} in \figref{fig:uml}.
Each primitive function except \acl{if},
whose calls are evaluated non-strictly
as described in \secref{sec:evaluation},
is implemented by
a private static method
\java{Primitive.exec\ijava{Prim}(Value)} or
\java{Primitive.exec\ijava{Prim}(Value,Value)}
(based on the function's arity),
where \ijava{Prim} is a Java ``version'' of the function's name;
the method returns a \java{Value}.
For instance, \java{Primitive.execCharCode(Value)} implements \acl{char-code}.
The package-private static method
\java{Primitive.call(Symbol,Value[])} evaluates a call
of the primitive function named by the \java{Symbol} argument
on the values in the \java{Value[]} argument,
by calling the appropriate \java{Primitive.exec\ijava{Prim}} method
and returning the result \java{Value}.
\java{Primitive} has no fields and only the above static methods;
no instances of this class are created.

The recognizers \acl{integerp}, \acl{consp}, etc.\
are implemented to return a \java{Symbol} for \acl{t} or \acl{nil},
based on whether the argument \java{Value} is an instance
of \java{Integer}, \java{Cons}, etc.

The destructors \acl{car}, \acl{numerator}, etc.\
are implemented to return information from the private fields.

The constructors \acl{complex} and \acl{cons}
are implemented via the factory methods
\java{Number.make} and \java{Cons.make}.
The constructor \acl{intern-in-package-of-symbol}
is implemented via the factory method \java{Symbol.make},
and it also calls the getter method \java{Symbol.getPackageName}
on the second argument.

The conversions \acl{char-code} and \acl{code-char}
are implemented by passing information from the private field of one class
to the factory method of the other class.
The conversion \acl{coerce}
has a slightly more laborious implementation,
which scans or builds a Java representation of an ACL2 list.

The arithmetic operation \acl{unary-{}-}
is implemented via package-private instance methods \java{negate()}
in the numeric classes.
\java{Primitive.execUnaryMinus},
when given \ijava{x} as argument,
calls \java{\ijava{x}.negate()}.
Dynamic dispatch selects a \java{negate} method
based on the runtime class of \ijava{x}.
\java{Integer.negate} calls \java{java.math.BigInteger.negate}
on its private field and uses the result to return a negated \java{Integer}.
Since $-(a/b) = (-a)/b$,
\java{Ratio.negate} calls \java{Integer.negate} on the numerator,
and uses the result to return a negated \java{Ratio}.
Since $-(a+bi) = (-a)+(-b)i$,
\java{ComplexRational.negate} calls \java{Rational.negate}
on the real and imaginary parts,
and uses the returned \java{Rational}s
to return a negated \java{ComplexRational};
each of these two calls to \java{Rational.negate} is, in turn,
dynamically dispatched to
\java{Integer.negate} or \java{Ratio.negate},
based on the runtime classes of the real and imaginary parts.

The arithmetic operation \acl{binary-{}+}
is similarly implemented via package-private instance methods \java{add(Value)}
in the numeric classes.
But the presence of the second argument
leads to a slightly more complicated interplay among the methods.
\java{Primitive.execBinaryPlus},
when given \ijava{x} and \ijava{y} as arguments,
calls \java{\ijava{x}.add(\ijava{y})},
dynamically dispatching based on the runtime class of \ijava{x}.
\java{Integer.add} splits into two cases:
if \ijava{y} is an \java{Integer},
an \java{Integer} sum is returned using
\java{java.math.BigInteger.add};
otherwise, the roles of \ijava{x} and \ijava{y} are swapped,
exploiting the commutativity of addition,
by calling \java{\ijava{y}.add(\ijava{x})},
which dynamically dispatches to a different \java{add} method.
\java{Ratio.add} performs an analogous split;
since $a/b+c/d = (ad+cb)/bd$,
this method calls the \java{multiply} methods,
further complicating the method interplay.
Since $(a+bi)+(c+di) = (a+c)+(b+d)i$,
\java{ComplexRational.add}
calls \java{Rational.add} on the real and imaginary parts,
which are further dynamically dispatched
to \java{Integer.add} or \java{Ratio.add}.

The arithmetic operations \acl{unary-{}/} and \acl{binary-{}*}
are implemented analogously to \acl{unary-{}-} and \acl{binary-{}+},
via \java{reciprocate()} and \java{times(Value)} methods
in the numeric classes.
There is some additional interplay among methods:
for instance, since $1/(a+bi) = (a/(a^2+b^2))-(b/(a^2+b^2))i$,
\java{ComplexRational.reciprocate} calls
all the arithmetic methods of \java{Rational}.

The arithmetic comparison \acl{<}
is implemented analogously to \acl{binary-{}+} and \acl{binary-{}*},
as part of an implementation of ACL2's total order
\acldoc{lexorder}{http://www.cs.utexas.edu/users/moore/acl2/manuals/current/manual/?topic=ACL2____LEXORDER}
via \java{compareTo(Value)} methods in \java{Value} and its subclasses,
which all implement
the \java{java.lang.Comparable<Value>} interface.
In the \java{compareTo} methods in the numeric classes,
when the roles of \ijava{x} and \ijava{y} are swapped
in the same way as in the \java{add} methods described above,
the result is negated before being returned,
because comparison, unlike addition and multiplication, is not commutative.

The equality \acl{equal}
is implemented via methods \java{equals(java.lang.Object)}
in \java{Value} and its subclasses,
which override \java{java.lang.Object.equals}.
These equality methods are implemented in the obvious way.

The implementation of \acl{bad-atom<=}
returns a \java{Symbol} for \acl{nil},
consistently with the raw Lisp code.

The functions \acl{pkg-imports} and \acl{pkg-witness}
are implemented as discussed in \secref{sec:environment}.
They throw an exception if the argument does not name a defined package,
matching ACL2's behavior.

All of these implementations do not check guards.
They handle \java{Value}s outside the guards
according to the applicable ACL2 completion axioms.


\subsection{Evaluation}
\label{sec:evaluation}

AIJ evaluates ACL2 terms via
(i) \java{eval(java.util.Map<Symbol,Value>)} methods
in \java{Term} and its subclasses,
and (ii) \java{apply(Value[])} methods
in \java{Function} and its subclasses.
This evaluation approach is well known \cite{lisp}.

The \java{eval} methods
take maps as arguments that bind values to variable symbols,
and return \java{Value} results.
\java{Constant.eval} returns the constant's value,
ignoring the map.
\java{Variable.eval} returns the value bound to the variable,
throwing an exception if the variable is unbound.
\java{Application.eval} recursively evaluates the argument terms
and then calls \java{Function.apply}
on the function and resulting values.
However, if the function represents \acl{if},
\java{Application.eval} first evaluates just the first argument,
and then, based on the result, either the second or third argument,
consistently with the non-strictness of \acl{if}.

The \java{apply} methods
take arrays of zero or more \java{Value}s as arguments,
and return \java{Value} results.
\java{LambdaExpression.apply}
evaluates the lambda expression's body
with a freshly created map that binds the values to the parameters---%
no old bindings are needed,
because lambda expressions are closed.
\java{NamedFunction.apply} calls
a public method \java{Environment.call(Symbol,Value[])}
with the name of the function and the argument values.
\java{Environment.call} operates as follows:
if the symbol names a primitive function,
it is forwarded, with the values, to \java{Primitive.call};
if the symbol names a function defined in the environment,
the lambda expression that defines the function
is applied to the values;
if the symbol does not name a primitive or defined function,
an exception is thrown.

AIJ evaluates ACL2 terms in a purely functional way,
without side effects.%
\footnote{Aside from exhausting the available memory,
which is, unavoidably, always a possibility.}
AIJ does not check the guards of primitive or defined functions.
The aforementioned method \java{Environment.call}
calls ACL2 functions on values only,
not on (names of) stobjs.


\subsection{Usage}
\label{sec:usage}

AIJ is designed to be used as follows by Java code outside AIJ's package:
\begin{enumerate}[nosep]
\item
Define all the ACL2 packages of interest
by repeatedly calling \java{Environment.addPackageDef}.
For each package,
use the factory methods of \java{PackageName} and \java{Symbol}
to build the name and the imported symbols.
Define both the built-in and user-defined packages,
in the order in which they appear in the ACL2 history.
This order ensures that \java{Symbol.make} does not throw an exception
due to an unknown package.
\item
Define all the ACL2 functions of interest
by repeatedly calling \java{Environment.addFunctionDef}.
For each function,
use the factory methods of the value and term classes
to build the name, the parameters, and the body.
The functions can be defined in any order,
so long as all the packages are defined before the functions (see step above).
\item
Call \java{Environment.setPackageWitnessName}
with the appropriate value from the ACL2 constant \acl{*pkg-witness-name*}.
\item
Call an ACL2 primitive or defined function as follows:
  \begin{enumerate}[nosep]
  \item
  Build the name of the ACL2 function to call,
  as well as zero or more ACL2 values to pass as arguments,
  via the factory methods of the value classes.
  \item
  Call \java{Environment.call} with
  the \java{Symbol} that names the function
  and the \java{Value} array of arguments.
  \item
  Unbuild the returned \java{Value} as needed to inspect and use it,
  using the getter methods of the value classes.
  \end{enumerate}
\item
Go back to step 4 as many times as needed.
\end{enumerate}

The above protocol explains why AIJ provides
a public API for unbuilding ACL2 values
but no public API to unbuild the other ACL2 entities (terms etc.).
The latter are built entirely by Java code outside AIJ's package,
which therefore has no need to unbuild the entities that it builds.
Values, instead,
may be built by executing ACL2 code that returns them as results:
Java code outside AIJ's package may need to unbuild the returned values
to inspect and use them.

Besides the structural constraints implicit in the Java classes,
and the existence of the referenced packages when building symbols
(necessary to resolve imported symbols),
AIJ does not enforce any well-formedness constraints
when building terms and other entities,
e.g.\ the constraint that
the number of arguments in a function call
matches the function's arity.
However, during evaluation, AIJ makes no well-formedness assumptions
and performs the necessary checks,
throwing informative exceptions if these checks fail.


\section{ATJ: The Code Generator}
\label{sec:atj}

ATJ is an ACL2 tool that provides an event macro
to generate Java code from specified ACL2 functions.
The generated Java code provides a public API to
(i) build an AIJ representation
of the ACL2 functions and other parts of the ACL2 environment
and (ii) evaluate calls of the functions on ACL2 values via AIJ.
The Java code generated by ATJ
automates steps 1, 2, and 3 in \secref{sec:usage}
and provides a light wrapper for step 4b,
while steps 4a and 4c must be still performed directly via AIJ's API.

ATJ consists of a few thousand lines of ACL2 code
(including blank lines, implementation-level documentation, and comments),
accompanied by a few hundred lines of user-level documentation in XDOC.
The implementation is thoroughly documented in XDOC as well.


\subsection{Overview}
\label{sec:gen-class}

ATJ generates a single Java file containing a single class,
with the following structure:
\begin{javablock}
  package \ijava{pname}; \cjava{// if specified by the user}
  import edu.kestrel.acl2.aij.*; \cjava{// all the AIJ classes}
  import ... \cjava{// a few classes of the Java standard library}
  public class \ijava{cname} \{ \cjava{// `ACL2' if not specified by the user}
      \cjava{// field to record if the ACL2 environment has been built or not:}
      private static boolean initialized = false;
      \cjava{// one method like this for each known ACL2 package:}
      private static void addPackageDef_\ijava{hex}(...) ...
      \cjava{// one method like this for each specified ACL2 function:}
      private static void addFunctionDef_\ijava{hex1}_\ijava{hex2}(...) ...
      \cjava{// API method to build the ACL2 environment:}
      public static void initialize() ...
      \cjava{// API method to evaluate ACL2 function calls:}
      public static Value call(Symbol function, Value[] arguments) ...
  \}
\end{javablock}
The file has the same name as the class;
it is (over)written in the current working directory,
unless the user specifies a directory.
ATJ directly generates Java concrete syntax,
via formatted printing to the ACL2 output channel associated to the file,
without going through a Java abstract syntax and pretty printer.


\subsection{Value and Term Building}
\label{sec:build-value-term}

As part of building an AIJ representation of the ACL2 environment,
the Java code generated by ATJ
builds AIJ representations of ACL2 values and terms:
function definitions include terms as bodies,
and constant terms include values.
It does so via the factory methods
discussed in \secrefII{sec:values}{sec:terms}.

In principle, ATJ could turn each ACL2 value or term
into a single Java expression with an ``isomorphic'' structure.
For example, the ACL2 value
\acl{((10 . $\!\!\!\!\!$\#\textbackslash{}A) . "x")}
could be built as follows:
\begin{javablock}
  Cons.make(Cons.make(Integer.make(10),
                      Character.make(65)),
            String.make("x"))
\end{javablock}

However, values and terms of even modest size (e.g.\ function bodies)
would lead to large expressions, which are not common in Java.
Thus, ATJ breaks them down into sub-expressions assigned to local variables.
For instance, the example value above is built as follows:
\begin{javablock}
  \cjava{// statements:}
  Value value1 = Integer.make(10);
  Value value2 = Character.make(65);
  Value value3 = Cons.make(value1, value2);
  Value value4 = String.make("x");
  \cjava{// expression:}
  Cons.make(value3, value4)
\end{javablock}

In general, ATJ turns each ACL2 value or term
into (i) zero or more Java statements that incrementally build parts of it
and (ii) one Java expression that builds the whole of it from the parts.
ATJ does so recursively: the expression for a sub-value or sub-term
is assigned to a new local variable that is used
in the expression for the containing super-value or super-term.
The top-level expressions are used
as explained in \secrefII{sec:build-pkgs}{sec:build-fns}.

To generate new local variable names,
ATJ keeps track of three numeric indices
(for values, terms, and lambda expressions---%
recall that the latter are mutually recursive with terms)
as it recursively traverses values and terms.
The appropriate index is appended to
`\java{value}', `\java{term}', or `\java{lambda}'
and then incremented.


\subsection{Package Definition Building}
\label{sec:build-pkgs}

The Java code generated by ATJ builds an AIJ definition
of each ACL2 package known when ATJ is invoked.
The names of the known packages are the keys of the alist
returned by the built-in function \acl{known-package-alist},
in reverse chronological order.

The AIJ definition of each of these packages
is built by a method \java{addPackageDef_\ijava{hex}}
(see \secref{sec:gen-class}),
where \ijava{hex} is an even-length sequence of hexadecimal digits
for the ASCII codes of the characters that form the package name.
For instance, the definition of the \acl{"ACL2"} package
is built by \java{addPackageDef_41434C32}.
This simple naming scheme ensures that the generated method names
are distinct and valid,
since ACL2 package names allow characters disallowed by Java method names.

Each \java{addPackageDef_\ijava{hex}} method
builds a Java list of all the symbols imported by the package,
which ATJ obtains via \acl{pkg-imports}.
Then the method calls \java{Environment.addPackageDef}
with the \java{PackageName} and the list of \java{Symbol}s.


\subsection{Function Definition Building}
\label{sec:build-fns}

The Java code generated by ATJ builds an AIJ definition
of each (non-primitive) ACL2 function specified via
one or more function symbols \sfn{1}, ..., \sfn{p} supplied to ATJ.
Each \sfn{i} implicitly specifies not only \sfn{i} itself,
but also all the functions called directly or indirectly by \sfn{i},
ensuring the ``closure'' of the generated Java code under ACL2 calls.

ATJ uses a worklist algorithm,
initialized with \acl{(\sfn{1} ... \sfn{p})},
to calculate a list of their closure under calls.
Each iteration removes the first function \iacl{fn} from the worklist,
adds it to the result list,
and extends the worklist with all the functions directly called by \iacl{fn}
that are not already in the result list.
Here `directly called by' means
`occurring in the \acl{unnormalized-body} property of';
occurrences in the guard of \iacl{fn} do not count,
because ATJ, like AIJ, ignores guards.
If \iacl{fn} has no \acl{unnormalized-body} property,
it must be primitive,
otherwise ATJ stops with an error---%
this happens if \iacl{fn} is a constrained, not defined, function.
If \iacl{fn} is in
\acl{logic-fns-with-raw-code} or \acl{program-fns-with-raw-code}
(see \secref{sec:evalsem}),
it must be in a whitelist of functions that are known to have no side effects;%
\footnote{This whitelist is currently a subset of \acl{logic-fns-with-raw-code}.
It consists of functions whose raw Lisp code makes them run faster
but is otherwise functionally equivalent to the ACL2 definitions.}
If \iacl{fn} has input or output stobjs,
ATJ stops with an error---%
this may only happen if \iacl{fn} is not primitive.

The AIJ definition of each of these functions
is built by a method \java{addFunctionDef_\ijava{hex1}_\ijava{hex2}}
(see \secref{sec:gen-class}),
where \ijava{hex1} and \ijava{hex2} are
even-length sequences of hexadecimal digits
for the ASCII codes of the package and symbol names of the function symbol.
For instance, the definition of the \acl{len} function
is built by \java{addFunctionDef_41434C32_4C454E}.
This simple naming scheme ensures that the generated method names
are distinct and valid,
since ACL2 package and symbol names allow characters
disallowed by Java method names.

Each \java{addFunctionDef_\ijava{hex1}_\ijava{hex2}} method
first builds a \java{Term}
from the \acl{unnormalized-body} property of the function,
as explained in \secref{sec:build-value-term}.
Then the top-level Java expression,
along with a \java{Symbol} for the function name
and with a \java{Variable} array for the function parameters,
is passed to \java{Environment.addFunctionDef}.


\subsection{Environment Building}
\label{sec:build-env}

The \java{initialize} method generated by ATJ (see \secref{sec:gen-class})
calls all the
\java{addPackageDef_\ijava{hex}} and
\java{addFunctionDef_\ijava{hex1}_\ijava{hex2}}
methods described in \secrefII{sec:build-pkgs}{sec:build-fns}.
The method also calls \acl{Environment.setPackageWitnessName}
with an argument derived from \acl{*pkg-witness-name*}.

The package definition methods are called
in the same order in which the corresponding packages are defined,
which is the reverse order of the alist returned by
\acl{known-package-alist}.

This ensures the success of the calls of \java{Symbol.make}
that build the elements of a package's import list.
For instance, if \acl{"P"} imports \acl{q::sym},
then \acl{"Q"} must be already defined when \acl{"P"} is being defined.
That is, \java{addPackageDef_51} must have already been called
when \java{addPackageDef_50} calls \java{Symbol.make("Q","SYM")}
as part of building \acl{"P"}'s import list,
which is needed to define \acl{"P"};
otherwise, \java{Symbol.make("Q","SYM")} would throw an exception
due to \acl{"Q"} being still undefined.

The function definition methods are called
after the package definition methods,
again to ensure the success of the \java{Symbol.make} calls.
The relative order of the function definitions is unimportant;
the result list returned by ATJ's worklist algorithm
(see \secref{sec:build-fns})
is in no particular order.

The \java{initialize} method may be called once by external code:
the method throws an exception
unless the \java{initialized} field (see \secref{sec:gen-class})
is \java{false},
and sets the field to \java{true} just before returning.


\subsection{Call Forwarding}
\label{sec:call-fwd}

The \java{\ijava{cname}.call} method generated by ATJ
(see \secref{sec:gen-class})
forwards the function name and the argument values
to \java{Environment.call},
after ensuring that the \java{initialized} field is \java{true},
i.e.\ that the ACL2 environment has been built.
It throws an exception if \java{initialized} is still \java{false}.


\section{Preliminary Tests and Optimizations}
\label{sec:tests}

The initial version of AIJ was deliberately written in a very simple way,
without regard to performance,
as a sort of ``executable specification'' in Java.
The reasons were to
increase assurance by reducing the chance of errors,
facilitate the potential verification of the code,
avoid premature optimizations,
and observe the impact of gradually introduced optimizations.

Performance has been tested mainly on three example programs.
The first is an ACL2 function that computes factorial non-tail-recursively.
The second is an ACL2 function that computes Fibonacci non-tail-recursively.
The third is a slightly modified version of
the verified ABNF grammar parser \cite{abnf}
from the ACL2 Community Books
\acldoc{abnf::grammar-parser}{http://www.cs.utexas.edu/users/moore/acl2/manuals/current/manual/?topic=ABNF____GRAMMAR-PARSER}:
the parser, in the \acl{:logic} part of \acl{mbe},
calls \acl{nat-list-fix} on its input list of natural numbers
just before reading each natural number,
which makes execution ``in the logic'' (which is how AIJ executes) unduly slow;
for testing AIJ more realistically,
the parser was tweaked to avoid these calls of \acl{nat-list-fix}.
The tweaked parser is about 2,000 lines (including blank lines),
including theorems to prove its termination so that it is in logic mode,
and including return type theorems to prove its guards.
The parser not only recognizes ABNF grammars,
but also returns parse trees.%
\footnote{\label{abnf-recognizer}%
Initially, tests were conducted
on a simplified version of the parser that only recognized ABNF grammars,
because ATJ did not support \acl{mbe},
which is used in the construction of parse trees
(defined via fixtypes
\acldoc{fty}{http://www.cs.utexas.edu/users/moore/acl2/manuals/current/manual/?topic=ACL2____FTY}).
After extending ATJ to support \acl{mbe},
testing was switched to the more realistic version of the parser
that also returns parse trees.}

Unsurprisingly, the initial version of AIJ was quite slow.
A re-examination of the code from a performance perspective
readily revealed several easy optimization opportunities,
which were carried out and are part of the current version of AIJ.
These are the main ones, in order:
\begin{enumerate}[nosep]
\item
The \java{Character} array representation of \java{String}s
was replaced with \java{java.lang.String}.
\item
\java{Symbol}s frequently used during evaluation,
such as the ones for
\acl{t}, \acl{nil}, and the names of the primitive functions,
were cached as constants instead of being built repeatedly.
\item
\java{Character}s were interned, as follows.
\java{Character} objects for all the 256 codes
were pre-created and stored into an array, in the order of their codes.
The factory method \java{Character.make(char)}
indexes the array with the input code and returns the corresponding object.
Since this ensures that there is just one object for each character code,
\java{Character.equals} uses pointer equality (i.e.\ \java{==})
and the default fast \java{java.lang.Object.hashCode} is inherited.
\item
\java{PackageName}s, \java{String}s, and \java{Symbol}s
were interned, similarly to \java{Character}s, as follows.
Since there is a potentially infinite number of them,
they are are created on demand.
For each of these three classes,
all the objects created thus far
are stored as values of a \java{java.util.Map},
whose keys are
\java{java.lang.String}s for \java{PackageName} and \java{String},
and \java{PackageName}s paired with \java{String}s for \java{Symbol}---%
the pairing is realized via nested maps.
Each factory method first consults the appropriate map,
either returning the existing object,
or creating a new one that is added to the map.
Similarly to \java{Character}'s interning,
the interning of these classes enables
the use of pointer equality in the equality methods
and the inheritance of the default fast hash code method.
\end{enumerate}
Thanks to AIJ's object-oriented encapsulation,
all these optimizations were easy and localized.
These optimizations did not involve ATJ,
because the code generated by ATJ is essentially used
just to initialize the ACL2 environment (see \secref{sec:build-env}),
which happens quickly
for the factorial and Fibonacci functions and for the ABNF parser.

Based on a few time measurements
on the ABNF parser and a few other artificial programs,
the above optimizations reduced execution time, very roughly,
by the following factors, one after the other:
2 for optimization \#1,
5 for optimization \#2, and
2 for optimizations \#3 and \#4---%
all combined, 20.

\tabrefIII{tab:fact}{tab:fib}{tab:abnf} report more systematic time measurements
for the factorial function, Fibonacci function, and ABNF parser.
Each row corresponds to an input of the program:
natural numbers for the factorial and Fibonacci functions;
ABNF grammars (all from Internet standards, including ABNF itself)
for the ABNF parser.
The `ACL2' columns are for execution in ACL2,
with guard checking (`g.c.') set to \acl{t},
i.e.\ typical execution,
and \acl{:none},
i.e.\ execution ``in the logic'';
the latter matches AIJ's execution.
The `AIJ' column is for execution with AIJ's current version.
Each cell contains minimum, average, and maximum real times from 10 runs,
in seconds rounded to the millisecond.
The ACL2 times were measured as the difference between the results of
\acl{read-run-time} just before and just after the call of
the factorial function, Fibonacci function, or top-level ABNF parsing function.
The Java times were measured as the difference between the results of
\java{java.lang.System.currentTimeMillis()}
just before and just after the call of
\java{\ijava{cname}.call} on the corresponding ACL2 function.
Given the AIJ evaluator's recursive implementation,
a larger stack size than the default must be passed to the JVM
(1 GB for these time measurements) to avoid a stack overflow.

The times in \tabref{tab:fact}
are all roughly comparable for each input,
with ACL2 faster on smaller inputs and AIJ faster on larger inputs:
presumably, most of the time is spent multiplying large numbers,
which all happens in \java{java.math.BigInteger} in the Java code
and in Lisp's bignum implementation in the ACL2 code,
dwarfing the contributions of ACL2 and AIJ proper,
especially for the larger inputs.%
\footnote{Even the initial, unoptimized version of AIJ took comparable times.}
The times in \tabref{tab:fib} differ:
looking at the averages,
AIJ is about 17--30 times slower than ACL2 with guard checking \acl{:none},
which is about 8--10 times slower than ACL2 with guard checking \acl{t}.
The times in \tabref{tab:abnf} differ as well:
looking at the averages,
AIJ is about 19--22 times slower than ACL2 with guard checking \acl{:none},
which is about 16--87 times slower than ACL2 with guard checking \acl{t};
nonetheless, the absolute times suggest that
the Java code of the parser is usable.%
\footnote{As another data point,
the simplified parser mentioned in \footref{abnf-recognizer}
was about 4--7 times faster than the current parser.}
Performance needs vary:
AIJ's current speed may be adequate to some applications,
such as security-critical interactive applications like cryptocurrency wallets.
Furthermore, as discussed in \secref{sec:future},
there are more opportunities to optimize AIJ.

\begin{table}[ht]
\centering
{\footnotesize
\begin{tabular}{|r||rrr|rrr|rrr|}
\hline
\multicolumn{1}{|c||}{\multirow{2}{*}{\textbf{Input}}} &
\multicolumn{3}{c|}{\textbf{ACL2\ \ \ [g.c.\ \acl{t}]}} &
\multicolumn{3}{c|}{\textbf{ACL2\ \ \ [g.c.\ \acl{:none}]}} &
\multicolumn{3}{c|}{\textbf{AIJ}} \\
& \multicolumn{1}{c}{\textbf{min}}
& \multicolumn{1}{c}{\textbf{avg}}
& \multicolumn{1}{c|}{\textbf{max}}
& \multicolumn{1}{c}{\textbf{min}}
& \multicolumn{1}{c}{\textbf{avg}}
& \multicolumn{1}{c|}{\textbf{max}}
& \multicolumn{1}{c}{\textbf{min}}
& \multicolumn{1}{c}{\textbf{avg}}
& \multicolumn{1}{c|}{\textbf{max}} \\
\hline\hline
1,000 &
0.000 & 0.000 & 0.001 & 
0.000 & 0.000 & 0.001 & 
0.003 & 0.005 & 0.012 \\ 
\hline
5,000 &
0.007 & 0.011 & 0.022 & 
0.007 & 0.009 & 0.011 & 
0.009 & 0.029 & 0.059 \\ 
\hline
10,000 &
0.031 & 0.035 & 0.038 & 
0.032 & 0.034 & 0.040 & 
0.026 & 0.035 & 0.068 \\ 
\hline
50,000 &
1.324 & 1.355 & 1.432 & 
1.319 & 1.328 & 1.337 & 
0.589 & 0.687 & 1.044 \\ 
\hline
100,000 &
6.280 & 6.385 & 6.604 & 
6.279 & 6.291 & 6.307 & 
2.340 & 2.547 & 2.705 \\ 
\hline
\end{tabular}
}
\caption{\label{tab:fact}
Time measurements for the factorial function.}
\end{table}

\begin{table}[ht]
\centering
{\footnotesize
\begin{tabular}{|r||rrr|rrr|rrr|}
\hline
\multicolumn{1}{|c||}{\multirow{2}{*}{\textbf{Input}}} &
\multicolumn{3}{c|}{\textbf{ACL2\ \ \ [g.c.\ \acl{t}]}} &
\multicolumn{3}{c|}{\textbf{ACL2\ \ \ [g.c.\ \acl{:none}]}} &
\multicolumn{3}{c|}{\textbf{AIJ}} \\
& \multicolumn{1}{c}{\textbf{min}}
& \multicolumn{1}{c}{\textbf{avg}}
& \multicolumn{1}{c|}{\textbf{max}}
& \multicolumn{1}{c}{\textbf{min}}
& \multicolumn{1}{c}{\textbf{avg}}
& \multicolumn{1}{c|}{\textbf{max}}
& \multicolumn{1}{c}{\textbf{min}}
& \multicolumn{1}{c}{\textbf{avg}}
& \multicolumn{1}{c|}{\textbf{max}} \\
\hline\hline
10 &
0.000 & 0.000 & 0.000 & 
0.000 & 0.000 & 0.000 & 
0.000 & 0.001 & 0.004 \\ 
\hline
20 &
0.000 & 0.000 & 0.000 & 
0.001 & 0.001 & 0.001 & 
0.019 & 0.030 & 0.053 \\ 
\hline
30 &
0.007 & 0.008 & 0.009 & 
0.061 & 0.063 & 0.079 & 
1.043 & 1.094 & 1.210 \\ 
\hline
40 &
0.727 & 0.734 & 0.750 & 
7.144 & 7.205 & 7.355 & 
126.167 & 127.149 & 129.959 \\ 
\hline
\end{tabular}
}
\caption{\label{tab:fib}
Time measurements for the Fibonacci function.}
\end{table}

\begin{table}[ht]
\centering
{\footnotesize
\begin{tabular}{|r||rrr|rrr|rrr|}
\hline
\multicolumn{1}{|c||}{\multirow{2}{*}{\textbf{Input}}} &
\multicolumn{3}{c|}{\textbf{ACL2\ \ \ [g.c.\ \acl{t}]}} &
\multicolumn{3}{c|}{\textbf{ACL2\ \ \ [g.c.\ \acl{:none}]}} &
\multicolumn{3}{c|}{\textbf{AIJ}} \\
& \multicolumn{1}{c}{\textbf{min}}
& \multicolumn{1}{c}{\textbf{avg}}
& \multicolumn{1}{c|}{\textbf{max}}
& \multicolumn{1}{c}{\textbf{min}}
& \multicolumn{1}{c}{\textbf{avg}}
& \multicolumn{1}{c|}{\textbf{max}}
& \multicolumn{1}{c}{\textbf{min}}
& \multicolumn{1}{c}{\textbf{avg}}
& \multicolumn{1}{c|}{\textbf{max}} \\
\hline\hline
ABNF grammar &
0.004 & 0.007 & 0.014 & 
0.109 & 0.113 & 0.117 & 
2.391 & 2.478 & 3.076 \\ 
\hline
JSON grammar &
0.001 & 0.002 & 0.006 & 
0.044 & 0.049 & 0.053 & 
1.011 & 1.023 & 1.031 \\ 
\hline
URI grammar &
0.002 & 0.003 & 0.006 & 
0.100 & 0.105 & 0.112 & 
2.218 & 2.233 & 2.251 \\ 
\hline
HTTP grammar &
0.002 & 0.004 & 0.010 & 
0.167 & 0.175 & 0.189 & 
3.577 & 3.597 & 3.626 \\ 
\hline
IMF grammar &
 0.009 &  0.014 &  0.021 & 
 1.028 &  1.079 &  1.414 & 
21.173 & 21.522 & 21.741 \\ 
\hline
SMTP grammar &
0.007 & 0.010 & 0.019 & 
0.398 & 0.404 & 0.411 & 
8.648 & 8.733 & 8.896 \\ 
\hline
IMAP grammar &
 0.020 &  0.026 &  0.030 & 
 2.198 &  2.267 &  2.481 & 
43.083 & 43.490 & 43.805 \\ 
\hline
\end{tabular}
}
\caption{\label{tab:abnf}
Time measurements for the ABNF parser.}
\end{table}

All the time measurements were taken on a
MacBook Pro (15-inch, 2017)
with 3.1 GHz Intel Core i7
and 16 GB 2133 MHz LPDDR3,
running macOS High Sierra Version 10.13.6.
The ACL2 times were measured with
commit 852ee0aca96deac2b3c062ee03f458acca668f6e from GitHub
running on 64-bit Clozure Common Lisp Version 1.11.5.
The Java times were measured with
the version of AIJ in the same commit from GitHub as above,
running on Oracle's 64-bit Java 10 2018-03-20,
Java SE Runtime Environment 18.3 (build 10+46).
Just before taking the measurements,
the machine was rebooted and only the necessary applications were started.

The performance of ATJ does not affect the performance of the Java code.
ATJ runs in 1--2 seconds on each of
the factorial function, Fibonacci function, and ABNF parser,
including the time to write the Java files;
this was measured by wrapping the calls of ATJ with \acl{time\$}.


\section{Future Work}
\label{sec:future}

Evaluating non-executable functions
(i.e.\ non-primitive and without an \acl{unnormalized-body} property),
by throwing an exception that mirrors the error that ACL2 yields,
is easy but not necessarily useful.%
\footnote{If support for evaluating non-executable function is added,
ATJ should still include an option to signal an error
when the worklist algorithm reaches a non-executable function.}
A planned extension is to support guards
and evaluation with different guard-checking settings,
in the same way as ACL2.
Support for functions with side effects will be added one at a time,
by writing native Java implementations (as done for the primitive functions)
that suitably mirror the ACL2 side effects in Java;
for instance, hard errors could be implemented as exceptions.
User-defined stobjs could be supported
by storing their contents in Java fields that are destructively updated;
since \java{state} is ``linked'' to external entities (e.g.\ the file system),
support for this built-in stobj will involve
the use of the Java API of those entities.
Supporting stobjs also involves extending AIJ's public API
to call ACL2 functions on stobj names, besides values.
Direct support for calling macros directly,
and for supplying named constants to function calls,
are also candidate extensions.

The generated method \java{\ijava{cname}.call}
described in \secref{sec:call-fwd}
does not provide much beyond calling \java{Environment.call} directly,
but is suggestive of additional functionality.
For example, future versions of \ijava{cname} could provide
a public method for each top-level target function \sfn{i} supplied to ATJ,
with no parameter for the function name,
and with as many \java{Value} parameters as the function's arity
instead of a single \java{Value} array parameter.
As another example, \ijava{cname} could provide
additional public methods to call each \sfn{i}
on objects of more specific types
(e.g.\ \java{Integer} instead of \java{Value}),
based on the guards.
The names of these methods should be derived
from the names of the corresponding functions,
according to safe but more readable schemes
than the one described in \secref{sec:build-fns}---%
in fact, a more readable scheme should be used
for the methods described in \secref{sec:build-pkgs} and \secref{sec:build-fns}
as well.

A reviewer suggested to
make the fields and methods of \java{Environment} non-static
and have multiple instances of this class at once.
This is worth exploring.

There are more optimization opportunities
beyond the ones already carried out and described in \secref{sec:tests}.
For example, now
each variable evaluation looks up the variable symbol
in the hash map that stores the current binding of values to variables.
As another example, now each function call
first looks up the function symbol in the hash map
that stores the function definitions in the environment,
and then, if no definition is found, it compares the function symbol
with all the primitive function symbols until a match is found.
Replacing or enhancing AIJ's representation of variable and function symbols
with numeric indices should make all these accesses much faster.%
\footnote{A preliminary experiment with just variable indices seems to
reduce execution times roughly by 2.}
As a third example, AIJ's evaluator could be re-implemented
as a loop with an explicit stack, instead of a recursion.
As a fourth example, many built-in ACL2 functions
could be implemented natively in Java (as done for the primitive functions),
instead of being interpreted.

A reviewer suggested to
implement \acl{hons}
\acldoc{hons-and-memoization}{http://www.cs.utexas.edu/users/moore/acl2/manuals/current/manual/?topic=ACL2____HONS-AND-MEMOIZATION}
in AIJ, and use it instead of \acl{cons}.
This amounts to interning all the Java objects that represent ACL2 values
(not just \java{Character}s, \java{String}s, and \java{Symbol}s---%
see \secref{sec:tests}),
enabling fast equality tests and hash code computations,
which could increase performance in some applications.
Perhaps future versions of AIJ and ATJ
could provide options to use \acl{hons} vs.\ \acl{cons}.

The eventual path to fast execution
is to avoid the interpretation overhead,
by having ATJ turn ACL2 functions into shallowly embedded Java representations,
as is customary in code generators.%
\footnote{Besides more conventional translation approaches,
a more speculative idea is
to generate the shallowly embedded representations
by partially evaluating \cite{parteval}
the AIJ interpreter on the deeply embedded representations
generated by ATJ.}
The shallow embedding will consist of Java methods
that implement the ACL2 functions,
with suitably matching signatures.
AIJ's representation of and operations on ACL2 values will still be used,
but AIJ's representation and evaluator of ACL2 terms will not.
Under certain conditions,
it should be possible to generate variants of these Java methods
that use more efficient representations and operations,
e.g.\ the Java \java{int} values and integer operations
when there is provably no wrap-around,
in particular leveraging ACL2's \acl{the} forms
and the associated guard verification,
which similarly help the Lisp compiler.
Given that generating these shallowly embedded representations
is inevitably more complicated and thus error-prone,
the slower but safer interpreted evaluation
could be still available as an option,
at least in the absence of verification.

The Java code generated by ATJ can be called by external Java code,
but not vice versa.
Allowing the other call direction may involve
suitable ACL2 stubs that correspond to the external code to be called.

The implementation of ATJ could be simplified
by directly generating a Java abstract syntax
and using a separable pretty printer to write abstract syntax to the file.

More ambitious projects are to
(i) verify the correctness of
AIJ's evaluator and primitive function implementations,
and (ii) extend ATJ to generate a proof of correctness
of the generated Java code,
like a verifying compiler.
Optimizing AIJ and generating shallowly embedded representations in ATJ
make these verification tasks harder;
an idea worth exploring is to perform a compositional verification
of optimized Java code against unoptimized Java code
and of the latter against ACL2 code.

The approach to generate Java code described in this paper,
including the envisioned extensions described in this section,
could be used to generate code in other programming languages.
In particular, the UML class diagram in \figref{fig:uml}
could be used for other object-oriented programming languages.


\section{Related Work}
\label{sec:related}

The author is not aware of any Java or other code generator for ACL2.

Several theorem provers (Isabelle, PVS, Coq, etc.)
include facilities to generate code in various programming languages
(Standard ML, Ocaml, Haskell, Scala, C, Scheme, etc.)
\cite{isa-codegen,pvs-codegen,coq-refman}.
These code generators use shallow-embedding approaches,
very different from ATJ and AIJ's deep-embedding approach.
These code generators may be more relevant
to future versions of ATJ and AIJ that use a shallow-embedding approach
(see \secref{sec:future}).
However,
the ACL2 language is quite different from the languages of those provers:
first-order vs.\ higher-order,
untyped with a fixed universe of (evaluation) values
vs.\ typed with user-definable types,
extra-logical guards vs.\ types that are part of the logic,
and so on.
Thus, only some of the ideas from those provers' code generators
may be relevant to ACL2.

As discussed in \secref{sec:motiv},
there are other ways for ACL2 code
to interoperate with code in other programming languages,
without the need for generating code in those programming languages from ACL2.
However, this obviated need should be balanced against
the issues with these approaches discussed in \secref{sec:motiv};
different approaches may be best suited to different applications.


\section*{Acknowledgements}

This work was supported by DARPA under Contract No.\ FA8750-15-C-0007.
Thanks to Matt Kaufmann
for useful discussions about the ACL2 evaluation semantics.
Thanks to the anonymous reviewers for valuable suggestions.


\bibliographystyle{eptcs}
\bibliography{paper}

\begin{thebibliography}{10}
\providecommand{\bibitemdeclare}[2]{}
\providecommand{\surnamestart}{}
\providecommand{\surnameend}{}
\providecommand{\urlprefix}{Available at }
\providecommand{\url}[1]{\texttt{#1}}
\providecommand{\href}[2]{\texttt{#2}}
\providecommand{\urlalt}[2]{\href{#1}{#2}}
\providecommand{\doi}[1]{doi:\urlalt{http://dx.doi.org/#1}{#1}}
\providecommand{\bibinfo}[2]{#2}

\bibitemdeclare{misc}{abcl}
\bibitem{abcl}
\emph{\bibinfo{title}{Armed Bear Common Lisp ({ABCL})}}.
\newblock \bibinfo{howpublished}{\url{https://abcl.org}}.

\bibitemdeclare{misc}{acl2-manual}
\bibitem{acl2-manual}
\emph{\bibinfo{title}{{ACL2} Theorem Prover and Community Books: User Manual}}.
\newblock
  \bibinfo{howpublished}{\url{http://www.cs.utexas.edu/~moore/acl2/manuals/current/manual}}.

\bibitemdeclare{misc}{apt}
\bibitem{apt}
\emph{\bibinfo{title}{{APT} ({Automated Program Transformations})}}.
\newblock
  \bibinfo{howpublished}{\url{http://www.kestrel.edu/home/projects/apt}}.

\bibitemdeclare{misc}{cffi}
\bibitem{cffi}
\emph{\bibinfo{title}{{CFFI}: The Common Foreign Function Interface}}.
\newblock \bibinfo{howpublished}{\url{https://common-lisp.net/project/cffi}}.

\bibitemdeclare{inproceedings}{abnf}
\bibitem{abnf}
\bibinfo{author}{Alessandro \surnamestart Coglio\surnameend}
  (\bibinfo{year}{2018}): \emph{\bibinfo{title}{A Formalization of the {ABNF}
  Notation and a Verified Parser of {ABNF} Grammars}}.
\newblock In: {\sl \bibinfo{booktitle}{Proc.\ 10th Working Conference on
  Verified Software: Theories, Tools, and Experiments ({VSTTE})}}.
\newblock \bibinfo{note}{To appear in Springer LNCS}.

\bibitemdeclare{inproceedings}{simplify}
\bibitem{simplify}
\bibinfo{author}{Alessandro \surnamestart Coglio\surnameend},
  \bibinfo{author}{Matt \surnamestart Kaufmann\surnameend} \&
  \bibinfo{author}{Eric \surnamestart Smith\surnameend} (\bibinfo{year}{2017}):
  \emph{\bibinfo{title}{A Versatile, Sound Tool for Simplifying Definitions}}.
\newblock In: {\sl \bibinfo{booktitle}{Proc.\ 14th International Workshop on
  the {ACL2} Theorem Prover and Its Applications ({ACL2-2017})}}, pp.
  \bibinfo{pages}{61--77}, \doi{10.4204/EPTCS.249.5}.

\bibitemdeclare{misc}{coq-refman}
\bibitem{coq-refman}
\emph{\bibinfo{title}{Coq 8.8.1 Reference Manual}}.
\newblock \bibinfo{howpublished}{\url{https://coq.inria.fr}}.

\bibitemdeclare{misc}{isa-codegen}
\bibitem{isa-codegen}
\bibinfo{author}{\surnamestart {Florian Haftmann with contributions from Lukas
  Bulwahn}\surnameend} (\bibinfo{year}{2017}): \emph{\bibinfo{title}{Code
  generation from Isabelle/HOL theories}}.
\newblock \bibinfo{howpublished}{\url{https://isabelle.in.tum.de}}.
\newblock \bibinfo{note}{Tutorial distributed with Isabelle/HOL}.

\bibitemdeclare{misc}{jni}
\bibitem{jni}
\emph{\bibinfo{title}{Java Native Interface {JNI} Specification}}.
\newblock
  \bibinfo{howpublished}{\url{https://docs.oracle.com/javase/10/docs/specs/jni}}.

\bibitemdeclare{book}{parteval}
\bibitem{parteval}
\bibinfo{author}{Neil~D. \surnamestart Jones\surnameend},
  \bibinfo{author}{Carsten~K. \surnamestart Gomard\surnameend} \&
  \bibinfo{author}{Peter \surnamestart Sestoft\surnameend}
  (\bibinfo{year}{1999}): \emph{\bibinfo{title}{Partial Evaluation and
  Automatic Program Generation}}.
\newblock \bibinfo{publisher}{Prentice Hall}.
\newblock \bibinfo{note}{\url{http://www.itu.dk/people/sestoft/pebook}}.

\bibitemdeclare{techreport}{acl2-logic}
\bibitem{acl2-logic}
\bibinfo{author}{Matt \surnamestart Kaufmann\surnameend} \&
  \bibinfo{author}{J~Strother \surnamestart Moore\surnameend}
  (\bibinfo{year}{1998}): \emph{\bibinfo{title}{A Precise Description of the
  {ACL2} Logic}}.
\newblock \bibinfo{type}{Technical Report}, \bibinfo{institution}{Department of
  Computer Sciences, University of Texas at Austin}.
\newblock
  \bibinfo{note}{\url{http://www.cs.utexas.edu/users/moore/publications/km97a.pdf}}.

\bibitemdeclare{article}{lisp}
\bibitem{lisp}
\bibinfo{author}{John \surnamestart McCarthy\surnameend}
  (\bibinfo{year}{1960}): \emph{\bibinfo{title}{Recursive Functions of Symbolic
  Expressions and Their Computation by Machine, {Part I}}}.
\newblock {\sl \bibinfo{journal}{Communications of the {ACM}}}
  \bibinfo{volume}{3}(\bibinfo{number}{4}), pp. \bibinfo{pages}{184--195},
  \doi{10.1145/367177.367199}.

\bibitemdeclare{inproceedings}{pvs-codegen}
\bibitem{pvs-codegen}
\bibinfo{author}{Nararajan \surnamestart Shankar\surnameend}
  (\bibinfo{year}{2017}): \emph{\bibinfo{title}{A Brief Introduction to the
  {PVS2C} Code Generator}}.
\newblock In: {\sl \bibinfo{booktitle}{Proc.\ Workshop on Automated Formal
  Methods ({AFM'17})}}.

\bibitemdeclare{book}{cltl2}
\bibitem{cltl2}
\bibinfo{author}{Guy~L. \surnamestart Steele\surnameend}
  (\bibinfo{year}{1990}): \emph{\bibinfo{title}{{Common Lisp} the Language}}.
\newblock \bibinfo{publisher}{Digital Press}.
\newblock
  \bibinfo{note}{\url{https://www.cs.cmu.edu/Groups/AI/html/cltl/cltl2.html}}.

\end{thebibliography}


\end{document}